\documentclass[]{article}

\usepackage{graphicx}
\usepackage{txfonts}

\newcommand{\be}{\begin{equation}}
\newcommand{\ee}{\end{equation}}
\newcommand{\ba}{\begin{eqnarray}}
\newcommand{\ea}{\end{eqnarray}}
\newcommand{\bc}{\begin{center}}
\newcommand{\ec}{\end{center}}

\begin{document}

\title{LS I +61 303 as a potential neutrino source
on the light of MAGIC results}

\author{Diego F. Torres$^{1,2}$ \& Francis Halzen$^3$}

\maketitle

 \noindent {\small {$^1$ Instituci\'o Catalana de Recerca i Estudis Avan\c{c}ats (ICREA)}\\
 {$^2$ Institut de Ci\`encies de l'Espai (IEEC-CSIC),
             Facultat de Ciencies,
              Universitat Aut\`onoma de Barcelona,
              Torre C5 Parell, 2a planta, 08193 Barcelona, Spain.
              Email: dtorres@ieec.uab.es} \\ {$^3$ Department of Physics, University
of Wisconsin, Madison, WI 53706}}

%\offprints{Diego F. Torres} \mail{dtorres@ieec.uab.es}

%\titlerunning{Neutrinos from LS I +61 303}
%\authorrunning{Torres \& Halzen}

%% abstract %%%%%%%%%%%%%%%%%%%%%%%%%%%%%%%%%%%%%%%%%%%%%%%%%

\abstract{ Very high energy $\gamma$-rays have recently been
detected from the microquasar LS I +61 303 using the MAGIC
telescope. A phenomenological study on the concomitant neutrinos
that would be radiated if the $\gamma$-ray emission is hadronic in
origin is herein presented. Neutrino oscillations are considered,
and the expected number of events in a km-scale
detector such as ICECUBE is computed under different assumptions
including orbital periodicity and modulation, as well as different precision in the 
modeling of the detector. We argue that the
upper limits already imposed on the neutrino emission of LS I +61 303 using AMANDA-II and the forthcoming
measurements by ICECUBE may significantly constrain -in an
independent and unbiased way- the $\gamma$-ray to neutrino flux    
ratio, and thus the possibility of a hadronic origin of the
$\gamma$-rays. The viability of hadronic models based on wind-jet
interactions in the LS + 61 303 system after MAGIC measurements is
discussed. }

%\keywords{gamma rays: observations, neutrinos, X-ray binaries:
%individual (LS I +61 303)}

\maketitle

\section{Introduction}

The possibility of an hadronic origin of high energy radiation from
microquasars and { gamma}-ray binaries have been recently extensively
discussed (e.g., see Romero et al. 2001, 2003, Bosch-Ramon et al.
2005 and references therein). 
{ Earlier works have already dealt with neutrino and high energy emission from galactic sources, and especially those presenting jets. Among them, Levinson and Blanford (1996) where among the first to suggest a possible correlation between microquasars and gamma-ray sources. Later,
Levinson and Waxman (2001) showed that if the energy content of the jets was dominated by an electron-proton plasma, a several hour outburst of 1--100 TeV neutrinos produced by photomeson interactions should precede the radio flares that are associated
with major ejections. Photopion production in jets was also considered by Distefano et al. (2002), by assuming the parameters of their model as inferred from radio observations. 
A recent review on the high energy aspects of astrophysical jets, together with more complete references to previous works, can be found in the work by Levinson (2006). }
Partially, recent work has been motivated by
the discovery of the presence of relativistic hadrons in microquasar
jets like those of SS 433, as inferred from iron X-ray line
observations (e.g. Migliari et al. 2002). In addition, some sources
(as LS 5039 and LS I +61 303) present quite inconspicuous behavior
at X-rays, with low variability and flux levels. Protons, generally
being subject to less efficient energy loss mechanisms than
electrons, could be accelerated to high energies generating only low
luminosity counterparts at other frequencies. Microquasars have then
been  proposed to be an additional source of cosmic rays (e.g.,
Heinz \& Sunayev 2002).
{ It is however true that the proof of protons being behind the highest energy photons detected so far has yet to be given at a level that not even reasonable doubt is tenable. At the moment, neither  the analysis of EGRET detections correlated with supernova remnants (see Torres et al. 2003 for a review) nor the recent HESS results (e.g., Aharonian et al. 2006) are conclusive in this respect. This is a central problem of cosmic-ray physics. However, key multiwavelength observations (for instance, for RX J0852.0-4622, see Aharonian et al. 2006b) are not far from elucidating the origin of the gamma radiation, at least for some cases. It is clear that the detection of concomitant neutrino emission from a hadronic source of gamma-rays would be a definite proof. }

Here, under the { model independent} assumption that the $\gamma$-ray spectrum measured
by the Major Atmospheric Gamma Imaging Cherenkov telescope (MAGIC,
Albert et al. 2006) results from inelastic $pp$ collisions, a lower
limit to the neutrino yield expected from LS I +61 303 is computed. { Disregarding the nature of the compact object in LS I +61 303, recently classified as a gamma-ray pulsar wind system by Dhawan et al. (2006), a hadronic interpretation of the gamma emission would result in a prediction of a neutrino yield.
In any case, if there are jets, their matter content is unknown. We note, however, that modeling the high energy radiation with $pp$ collisions is not discarding leptonic contributions (see e.g., Bosch-Ramon et al. 2006), since those might dominate at lower gamma-ray energies and at other locations in the orbit of the binary system, where electron losses are diminished. The appeal of hadronic emission of TeV radiation in close X-ray binary systems can be understood if one considers the losses and acceleration process for electrons, under the assumptions of a given magnetic field. We discuss these issues in more detail below. In this paper, we}
discuss how this $pp$ assumption can be tested using neutrino
telescopes under different scenarios for variability, and whether
earlier proposed hadronic microquasar models can still comply with
current observational constraints. The key approach we propose here
is that an experiment such as ICECUBE could test the level of
enhancement of neutrino emission as compared with that detected at
TeV $\gamma$-ray energies. By comparing with theoretical modeling of
photon absorption, ICECUBE can ultimately test or disprove a
hadronic scenario for LS I +61 303.

\section{LS I +61 303: basic facts}

LS I +61 303 shares with LS 5039 the quality of being the only two
known microquasars that are spatially coincident with sources above
100 MeV listed in the Third Energetic Gamma-Ray Experiment (EGRET)
catalog (Hartman et al. 1999), and the only two detected so far at
higher $\gamma$-ray energies (Aharonian et al. 2005, Albert et al.
2006). These sources both show low X-ray emission and variability.
Optical spectroscopic observations of LS I +61 303, which is located
at a distance of 2.0$\pm$0.2 kpc (Frail \& Hjellming 1991), revealed
a rapidly rotating B0 main sequence star with a stable shell and
radial velocities compatible with binary motion in a 26.5 day
orbital period (Hutchings \& Crampton 1981). The most accurate value
of the orbital period, $P_{\rm orb}$=26.4960$\pm$0.0028 d, comes
from the analysis of more than 20 years of periodic radio outbursts
(Gregory et al. 2002). The maximum of the radio outbursts varies
between phase 0.45 and 0.95 (assuming $T_0$=JD 2.443.366,775)
following a modulation of 4.6 years. An X-ray outburst starting
around phase 0.4 and lasting up to phase 0.6 has also been detected
(Goldoni \& Mereghetti 1995, Taylor et al. 1996, Harrison et al.
2000). The periastron takes place at phase 0.23 and the eccentricity
is 0.72$\pm$0.15. { Extended jet-like and apparently
precessing radio emitting structures at angular extensions of
0.01-0.05 arcsec have been reported by Massi et al. (2001, 2004);
this discovery has supported the microquasar interpretation of LS I
+61 303. The 
uncertainty as to what kind of
compact object, a black hole or a neutron star, is part of the
system (e.g., Casares et al. 2005), seems to have been lifted while this paper was under review by Dhawan et al. (2006). These authors have presented 
preliminary results from a July 2006 VLBI campaign in which 
rapid changes are seen in the orientation of a cometary tail at periastron.
This tail is consistent with it being the result of a pulsar wind (in which case, the gamma-emission should be the result of the interaction of this wind with the companion star outflow). 
% The length of
% the tail (6.5mas) corresponds to a stellar wind at least v~7500 km/s (assuming 3 days % to make tail).  At other phases the tail is inconspicuous, the shift of centroid from day % to day gives v ~1800 km/s max, <1000km/s typical.
% For comparison, V_max of the pulsar in the orbit is ~300km/s
No large features or higher velocities were noted on any of the observing days, i.e., there was no notice of the high 0.6$c$ velocity outflow as reported by
Massi et al. 2005, which implies at the least its non-permanent nature. The changes within 3 hours were found to be insignificant, so the
velocity can not be much over $0.05\,c$. 
In fact, tail velocities around 7500 km s$^{-1}$ were measured at periastron.
If confirmed, these results would classify LS I +61 303 as a pulsar wind system (a gamma-ray binary) instead of a microquasar (powered by mass accretion). Debate is alive.}

\section{MAGIC detection of LS I +61 303}

The spectrum derived from MAGIC data between ~200 GeV and ~4 TeV
at orbital phases between 0.4 and 0.7 is fitted by a power law
function: \be F_\gamma = (2.7 \pm 0.4 \pm 0.8) \times 10^{-12}
(E/{\rm TeV})^{-2.6 \pm 0.2 \pm 0.2} \; {\rm cm}^{-2} {\rm s}^{-1}
{\rm  TeV}^{-1}, \ee with the errors quoted being statistical and
systematic, respectively (Albert et al. 2006). MAGIC measurements
showed that the very high energy $\gamma$-ray emission from LS I
+61 303 is variable. The maximum flux corresponded to about 16\%
of that of the Crab Nebula, and was detected around phase 0.6 with
8.7$\sigma$ of significance. In addition, since the system was
observed in different orbital periods and the detections occurred
at similar orbital phases, there is a hint (although not yet a
{ published} proof) of a periodic nature of the VHE $\gamma$-ray emission. A
definitive proof or disproof of periodicity of the $\gamma$-ray
emission in LS I +61 303 awaits { the publication of } further MAGIC observations.

\section{Timescales and models}

{

When analyzing the possibility of a leptonic origin of the gamma-ray emission (in such a case the neutrino production would be quenched), the location of the emitter within the source and the dichotomy between a fast jet or a slower cometary tail wind are crucial. To understand why, one can compare the relevant timescales and impose constraints upon the magnetic field. 
The maximum energy of accelerated electrons is generally computed by requesting that the acceleration timescale equals that of the losses (a.k.a. cooling), where the latter is provided by the electron's radiation via synchrotron and Compton mechanisms.  The acceleration timescale is (e.g., Malkov \& Dury 2001, Aharonian et al. 2005)
\be t_{\rm acc}=\eta\, \frac{r_{\rm L}}{c} 
\approx 0.11\, E_{\rm TeV}\, B_{\rm G}^{-1}\, \eta \ \rm s , \ee
where $r_{\rm L}=E/eB$ is the Larmor radius,
$E_{\rm TeV}=E/1\; \rm TeV$, 
and $B_{\rm G}=B/1 \; \rm G$ is the strength of the ambient 
magnetic field. $\eta$ gives account of the efficiency of acceleration: in the case of 
extreme accelerators (maximum possible acceleration rate
allowed by classical electrodynamics) $\eta \rightarrow 1,$ whereas for  
shock acceleration in the Bohm diffusion regime 
$\eta \approx 10 (v/c)^{-2}$ (see e.g. Malkov \& Dury 2001). 
For the Compton scattering we consider both, the Thompson and the Klein-Nishina
regimes. In the former (say, for $E_e \ll 0.1$ TeV) the cooling time is inversely proportional to the electron energy, 
\be t_{\rm Thom}\approx 0.03 E_{\rm TeV}^{-1} \ \rm s  . \ee
In the latter, the characteristic Compton cooling timescale 
can be approximated by 
\be t_{\rm KN} \approx 34  
w_0^{-1} E_{\rm TeV}^{0.7} \ \rm s , \ee 
where $w_0= w_{\rm r}/500~{\rm erg} 
\ {\rm cm}^{-3}$. The Compton scattering of TeV electrons against
starlight occurs in the Klein-Nishina regime. By equating the acceleration and
Compton timescales, $t_{\rm acc} = t_{\rm KN}$, we obtain 
\begin{equation}
E_{\rm e,max} \simeq  82657.9
[B_{\rm G} (v/c)^2  
w_0^{-1}]^{3.3}~{\rm TeV} \simeq 2 [B_{\rm G} (v/0.2 c)^2  
w_0^{-1}]^{3.3},
\label{Emax1}
\end{equation}
where the first equality is obtained just replacing the previous formulae and doing the algebra, and the second equality provides a useful scaling especially for microquasar jets, where velocities of 0.2$c$ are tenable (e.g., as in LS 5039).
The companion star of the compact object in LS I +61 303 is a BO V with a temperature about 22500 K. Its optical luminosity is taken to be $L_\star \approx 2 \times 10^{38} \ \rm erg/s$, and periastron and apastron are reached at about 2.5 and 14.5 stellar radii (e.g., Romero et al. 2005 and references therein). Assuming that $R_\star \simeq 10 R_\odot$, as appropriate for a Bo V/Be star, periastron (apastron) occurs at 
$1.7 \times 10^{12} \ \rm cm$ ($1.0 \times 10^{13} \ \rm cm$). The energy density of the starlight \be w_r=\frac{L_\star}{4 \pi R^2 c} , \ee varies then between 200 and 5 erg cm$^{-3}$.
Formally,  then, even for slow winds, a magnetic field $B > 1$~G allows for the maximum energy of accelerated electrons to approach/exceed 10 TeV. For such fields however, synchrotron emission also plays a role. The synchrotron timescale is
\be t_{\rm sync} \approx 400 B_G^{-2} E_{\rm TeV}^{-1} \ \rm s, \ee
that is, for instance, at $B_{\rm G}\sim 10$, the cooling is dominated by synchrotron radiation rather than Compton processes.
Together with the equality $t_{\rm acc} = t_{\rm sync}$, gives
\begin{equation} 
E_{\rm e,max}  \approx 19 B_{\rm G}^{-1/2} \, (v/c)~{\rm TeV} \ .
\label{Emax2}
\end{equation}
These equations imply that  
electrons could be accelerated to energies exceeding 
10 TeV in a low-magnetic field environment and/or low stellar density $w_0$. 
In the case of a jet system, low magnetic fields can be attained far from the inner parts of the jets, out of the core of the binary system. A periodic feature in the gamma-radiation
(which is evidence of $\gamma$-ray 
production inside the binary system)
would then likely favor hadrons, which cool less efficiently, as the primary particles. 
Their maximum  energy would be determined by 
the condition $r_L \leq R_{\rm jet}$, which  gives 
\be E_p \leq 3 \times 10^{15} 
(R_{\rm jet}/10^8~{\rm cm})(B/10^5~{\rm G})  \ \rm eV, \ee i.e. and inner-jet magnetic field of about $10^5$ G can produce PeV primaries. These protons are not sufficiently energetic to interact with starlight to trigger photomeson processes, and since the X-ray emission from LS I +61 303 is not indicative of the existence of an important accretion disk, the photomeson process against the X-ray photons is expected to be sub-dominant as well (the X-ray luminosity is only about 10$^{33}$ erg s$^{-1}$). % Goldoni - Mereghetti, us in Como with Chandra.
It is also possible to see that pion production would inevitably lead to neutrino emission (otherwise stated, that pions decay before interacting in the media), and also that the production of neutrinos from the subsequent muon decay
also proceed with high probability as long as the magnetic 
field does not exceed $B \sim 10^6$~G (see the computations from Aharonian et al. 2005 for the case of LS 5039: since LS I +61 303 have similar X-ray fluxes, they can be adapted vis a vis to this scenario).

%FROM AHARONIAN ET AL: In the lab frame, the decay time of  
%charged pions responsible for TeV neutrino production is 
%$t_{\pi^\pm}= (E_\pi/ m_\pi c^2)\,\tau_{\pi^\pm} 
%\approx 2.5 \times 10^{-3} (E_\pi/10 \ \rm TeV)$~s.  
%On the other hand, the cooling time  of $\pi^\pm$ 
%due to inelastic $\pi p$ and $\pi \gamma$ 
%interactions depends on the ambient gas $n_p$ 
%and photon $n_x$ densities:  
%$t_{\pi p} \sim 10^{-3} (n_p/10^{17}~{\rm cm^{-3}})^{-1}$ s, 
%and 
%$t_{\rm \pi \gamma} \sim 5 \times 10^{-3} 
%(n_X/10^{20} {\rm cm^{-3}})^{-1}$ s.
%For typical parameters characterizing LS~5039, 
%the number density of $X$-ray  
%photons and the plasma density
%in the region  $R \geq 10^7$~cm do not exceed 
%$10^{20} \ \rm cm^{-3}$ and 
%$10^{17} \ \rm cm^{-3}$, respectively.   
%Therefore charged pions decay to $\mu$ and $\nu_\mu$
%before interacting with the ambient photons and protons. 
%The production of neutrinos from the subsequent muon decay
%also proceed with high probability as long as the magnetic 
%field does not exceed $B \sim 10^6$~G. This follows 
%directly from the 
%comparison of the decay time of muons,  
%$t_\mu= (E_\mu/m_\mu c^2)\tau_{\mu} \simeq 
%0.2(E_\mu/10 \rm TeV)$~s, with their
%synchrotron cooling time
%$t_{\rm sy} \approx 0.07 
%\,(B/10^6 \rm G)^{-2} (E_\mu/10~{\rm TeV})^{-1}$~s.

Following the recent results form Dhawan et al. (2006), LS I +61 303 would be similar to the system PSR 1259-63/SS2883. In the latter, a 48 ms radio pulsar is found in an eccentric orbit around a 10th magnitude main sequence star with an strong wind, a Be star with $T_{\rm eff} \sim 27 000 $ K. LS I +61 303 would have, however, a much shorter period. The PSR 1259-63 system has been observed by CANGAROO (Kawachi et al. 2004) and HESS (Aharonian et al. 2005) and modeled by both, a mainly leptonic (e.g., Kirk et al. 1999) and a mainly hadronic contribution (Kawachi et al. 2004). 
Interestingly,the observed lightcurve of the PSR 1259-63 system seems to be qualitatively similar to the prediction made by the model of Kawachi 
et al. (2004) where hadronic interactions and neutral pion production in the
mis-aligned stellar disk plays a dominant role in the gamma-ray production mechanism.
Neronov et al. (2006) have also presented a hadronic interpretation of the PSR 1259-63  HESS data, where the TeV emission is the result of $pp$ collisions between protons accelerated in the pulsar wind and nuclei resident in the wind of the stellar companion.
Albeit a full hadronic model of LS I +61 303, in the case it actually is a pulsar wind system, is still missing and is beyond the scope of the present work, one certain sub-product of a hadronic gamma-ray production would be the concomitant emission of neutrinos. In what follows, we analyze in a model independent way --starting from the MAGIC data-- up to what level a detector like ICECUBE could detect such neutrinos.\footnote{ { NEMO/KM3NET, having a planned effective area similar in size to ICECUBE --see, e.g., Distefano et al. 2006 for a general discussion-- would also be a key player in the possible detection of this system. However, due to location in the southern hemisphere, ICECUBE is better suited for an observational study on  LS I +61 303 and we focus the discussion on it. These alternative instruments would of course be key for the detection of southern hemisphere sources, such as LS 5039.} }

}

\section{Neutrino yields}

\subsection {Neutrino fluxes}

Inelastic $pp$ collisions lead to charged (2/3) and neutral (1/3)
pions. Thus, if hadronic interaction are involved in the production
of the $\gamma$-ray flux, a comparable emission of $\nu_\mu$ and
$\nu_e$ is to be expected. For a detailed discussion of the
computations of $pp$ collision $\gamma$-rays see, for example, the
appendices in the works by Torres (2004) and Domingo-Santamar\'ia \&
Torres (2005).
The neutrino flux at the production site can be obtained from the
$\gamma$-ray flux measured at Earth, although with the caveat that
the latter can be affected by absorption. Thus, the so-predicted
neutrino flux is to be considered as a lower limit to the one
actually produced. In the GeV and TeV regime, if $\gamma$-rays are
significantly absorbed, the $\gamma$-ray spectrum may be steeper
than the neutrino one, what further secures the lower limit quality
of the estimation of the neutrino flux when starting from that in
$\gamma$-rays. The $\nu$-flux is $ {F_\nu^i}^0=k^i F_\gamma
\label{1} $ where  $k^i=(Q_\nu^i/Q_\gamma)$ stands for the ratio of
-at the source- emissivities  of the particular kind of neutrinos
$i=\mu,e,\tau$ or their antineutrinos, and $\gamma$-rays. This ratio
depends on the value of the photon spectral index, which (before
absorption of $\gamma$-rays proceeds) is to be shared by the
neutrino spectrum. The most recent derivation of $k^i$ has been
presented by Cavassinni et al. (2006), using the numerical package
{\sc PYTHIA}. Details of the prescription we use below and the way
in which it was derived can be found there. Galactic distances are
much larger than the neutrino oscillation length for energies around
TeV. Then, flavor oscillation probabilities have to be considered in
computing the neutrino flux at Earth. After propagation in vacuum,
the original neutrino flux will be modified according to $ F_i  =
\sum_{j = e,\mu,\tau}P_{ij}F_j^0 , $ where $F_j^0$ is the expected
neutrino flux in the absence of oscillations, and $P_{ij}$ the
probabilities of inter-conversion among neutrino flavors (Cavassinni
et al. 2006). Assuming that the CP violating phase is negligible { (e.g., 
Constantini and Vissani 2005)},
oscillations almost completely isotropize the signal, and generate a
non-negligible $\nu_\tau$ flux at Earth, even when it is negligible
at the source.

Taking into account the measured (average) MAGIC spectrum of
$\gamma$-rays, the predicted neutrino fluxes at the source out of
this spectrum, following the previously commented derivation, are
\begin{equation}
\left(%
\begin{array}{c}
        F_{\nu_\mu }^0\\
  F_{{\bar \nu}_\mu}^0 \\
           F_{\nu_e}^0 \\
    F_{{\bar \nu}_e}^0 \\
\end{array}%
\right) = \left(%
\begin{array}{c}
  9.72\times 10^{-13} \\
  9.72\times 10^{-13} \\
  5.40\times 10^{-13} \\
  4.85\times 10^{-13} \\
\end{array}%
\right)  \left( \frac{E_\nu}{\rm TeV}\right) ^{-2.6}  {\rm TeV}^{-1} {\rm cm}^{-2} {\rm s}^{-1},
\end{equation}
whereas the fluxes at Earth after taking into account $\nu$-oscillations are
\begin{equation}
\left(%
\begin{array}{c}
        F_{\nu_\mu }\\
  F_{{\bar \nu}_\mu} \\
           F_{\nu_e} \\
    F_{{\bar \nu}_e} \\
F_{\nu_\tau }\\
 F_{{\bar \nu}_\tau} \\
\end{array}%
\right) = \left(%
\begin{array}{c}
 4.92\times 10^{-13} \\
  4.81\times 10^{-13} \\
  5.11\times 10^{-13} \\
  4.81\times 10^{-13} \\
4.93\times 10^{-13} \\
  4.92\times 10^{-13} \\
\end{array}%
\right) \left( \frac{E_\nu}{\rm TeV}\right) ^{-2.6}  {\rm TeV}^{-1} {\rm cm}^{-2} {\rm s}^{-1}.
\label{nu}
\end{equation}
Errors in the $\gamma$-ray spectrum (about 30\% in the
normalization and 10\% in the slope, including statistics and
systematics in quadrature) would propagate directly into the
neutrino flux. It is to be stressed that the latter neutrino
fluxes are lower limits only. Since absorption of $\gamma$-rays in
the stellar photon fields of the binary system is unavoidable, in
the case of a hadronic production of $\gamma$-rays, the neutrino
flux ought to be larger than the $\gamma$-ray one (e.g., Aharonian
et al. 2005, Christiansen et al. 2006).
In the case of LS 5039, for instance, Aharonian et al. (2005) studied the
potential of such source as a neutrino emitter after the measurement
by HESS, and proposed that the neutrino flux expected above 1 TeV
could be up to a hundred times larger than the $\gamma$-ray flux
detected in that energy band. In such an increased range, ANTARES,
an experiment with similar effective area for neutrino detection
than AMANDA (see below), would be able to test the putative neutrino
emission. In the case of LS I +61 303, and in the framework of the
hadronic microquasar model developed by Romero et al. (2003) and
Torres et al. (2005), Christiansen et al. (2006) reported that the
opacity-corrected $\gamma$-ray flux above 350 GeV dropped from $7.1
\times 10^{-11}$ cm$^{-2}$ s$^{-1}$ to $1.4 \times 10^{-12}$
cm$^{-2}$ s$^{-1}$, at the periastron of the system, i.e., a
reduction by a factor of 50 in $\gamma$-ray flux, that would not
affect the associated neutrino production (we discuss further this
model below). Neutrino telescopes can constraint this very factor of
enhancement, the neutrino to photon ratio.

\subsection{Testing emission scenarios with neutrinos}

Indeed, we argue that the combination of the $\gamma$-ray flux
measured by MAGIC from LS I +61 303 with the existing upper limits
from neutrino experiments already restricts the enhancement factor,
and thus provides constraints for a hadronic origin of the
$\gamma$-ray radiation from LS I +61 303. We first note that the
neutrino flux of Eq. (\ref{nu}) is consistent with AMANDA-II upper
limits on this source (Ackermann et al. 2005). The latter has been
imposed at 90\% CL as $2.4 \times 10^{-8}$ cm$^{-2}$ s$^{-1}$,
integrated for energies above 10 GeV and assuming an spectral slope
of $-2$. The quoted limit correspond to the 2000-2002 data sample.
This upper limit is not very restrictive, although it is only a
factor of  25 larger (disregarding the possible difference that the
slope used to obtain it may introduce, which does not affect the
concept) than the flux given in Eq. (2). Ackerman has recently
presented a new limit from 4 years of AMANDA II data, at a level of
$6 \times 10^{-9}$ cm$^{-2}$ s$^{-1}$. The combined $ F_{\nu_\mu }$
and $F_{{\bar \nu}_\mu} $ lower limit flux yielded by LS I +61 303
at Earth as deduced in Eq. (\ref{nu}), when integrated above 10 GeV,
is only about a factor of 6 smaller than the improved AMANDA-II
upper limit.
%Assuming that the differences in slopes between the
%MAGIC measurement and the the one chosen to present the AMANDA-II
%upper limits introduces a negligible difference,
Then, we can compute how much time the next generation of neutrino
experiments would need to constrain further a possible hadronic
emission; i.e., how well can neutrino experiments constrain the
neutrino to photon ratio.\footnote{If the neutrino spectrum is
harder than the photon one (as assumed for instance in the case of
LS 5039 by Aharonian et al. 2005), the neutrino flux would be even
larger than just the scaling up of the $\gamma$-ray flux, so that to
impose a constraint over the neutrino to $\gamma$-ray ratio, the
most conservative choice is that the change in spectrum is not
significant, which is an assumption we follow. For instance, if the
increased neutrino flux is due to absorption, the neutrino flux
would be similar to the one we use but with a standard $E^{-2}$
spectrum (see, e.g., Halzen \& Hooper 2005). This will increase the
number of events.} To do this, we briefly summarize how to
approximately compute the signal and background events for an
ICECUBE-like detector.

\subsubsection{Neutrino detection}

Neutrino telescopes search for up-going muons produced deep in the
Earth, and are mainly sensitive to the incoming flux of $\nu_\mu$
and ${\bar \nu}_\mu$.  ICECUBE  will consist of 4800
photomultipliers, arranged on 80 strings placed at depths between
1400 and 2400 m under the South Pole ice (e.g., Halzen 2006). The
strings will be located in a regular space grid covering a surface
area of 1 km$^2$. Each string will have 60 optical modules (OM)
spaced 17 m apart. The number of OMs which have seen at least one
photon (from \v{C}erenkov radiation produced by the muon which
resulted from the interaction of the incoming $\nu$ in the earth and
ice crust) is called the channel multiplicity, $N_{\rm ch}$. The
multiplicity threshold is set to $N_{\rm ch}=10$, which corresponds
to an energy threshold of 200~GeV, the same threshold of the MAGIC
$\gamma$-ray observations. The angular resolution of ICECUBE will be
$\sim 0.7^\circ$.
A first estimation of the event rate of the atmospheric
$\nu$-background that will be detected in the search bin is given
by (e.g., Anchordoqui et al. 2003, Romero \& Torres 2003)
\begin{equation}
\left. \frac{dN}{dt}\right|_{\rm B} = A_{\rm eff}\, \int dE_\nu
\,\frac{d\Phi_{\rm B}}{dE_\nu}\, P_{\nu \to \mu}(E_\nu)\,\,\Delta
\Omega\,, \label{background}
\end{equation}
where $A_{\rm eff}$ is the effective area of the detector, $\Delta
\Omega \approx 1.5 \times 10^{-4}$~sr is the angular size of the
search bin, and $d\Phi_{\rm B}/dE_\nu \lesssim 0.2 \,(E_\nu/{\rm
GeV})^{-3.21}$~GeV$^{-1}$ cm$^{-2}$ s$^{-1}$ sr$^{-1}$  is the
$\nu_\mu + \bar \nu_\mu$ atmospheric $\nu$-flux~(Volkova 1980,
Lipari 1993). Here, $P_{\nu \to \mu} (E_\nu)$ denotes the
probability that a $\nu$ of energy $E_\nu$ on a trajectory through
the detector, produces a muon.  For $E_\nu \sim 1 - 10^3~{\rm
GeV}$, this probability is $ \approx 3.3 \times 10^{-13}
\,(E_\nu/{\rm GeV)}^{2.2}$, whereas for $E_\nu
> 1~{\rm TeV}$,
$P_{\nu \to \mu} (E_\nu) \approx 1.3 \times 10^{-6} \,(E_\nu/{\rm
TeV)}^{0.8}$ ~(Gaisser et al. 1995).
On the other hand, the $\nu$-signal is similarly obtained as
\begin{equation}
\left. \frac{dN}{dt}\right|_{\rm S} =  A_{\rm eff} \,\int dE_\nu
\,  ( F_{\nu_\mu }+F_{{\bar \nu}_\mu} ) \,P_{\nu \to \mu}(E_\nu)\
\,, \label{yellowsubmarine}
\end{equation}
where $( F_{\nu_\mu }+F_{{\bar \nu}_\mu} )$ is the incoming
$\nu_\mu$-flux.
{ In the previous integrals we use both expressions for $P_{\nu \to \mu} (E_\nu)$
according to the energy, and integrate from 200 GeV up to 10 TeV. The approximate validity of these expressions for the probability can be indirectly verified by the comparison of the neutrino background estimates and measurements in experiments such as ICECUBE.}

\begin{figure}[t]
\centering
\includegraphics[height=8cm, width=7.cm]{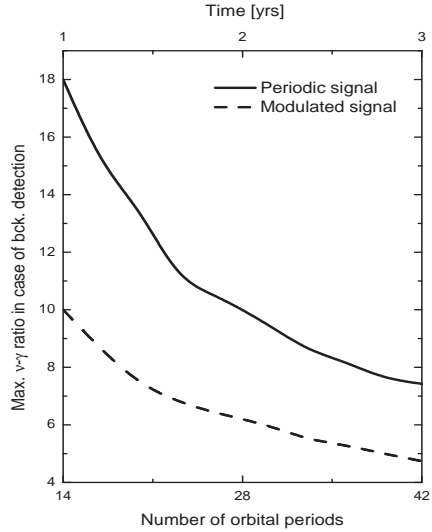}
\caption{Maximum neutrino to photon ratio compatible with the 95\%
CL interval for the corresponding neutrino background in the given
integration time. See text for a more detailed explanation. }
\label{1}
\end{figure}

\begin{table} %[t]
\centering \caption{Computed neutrino yields for two years of
observations in ICECUBE assuming the neutrino to photon ratio is
equal to 1.}
\begin{tabular}{lll}

\hline
Orbital  scenario & Signal Events  & Background Events \\
\hline

Strict Periodicity & 0.6 & 5.0 \\
Modulation & 1.3 & 10.0 \\

\hline
%\multicolumn{3}{l} {$^1$$R_{\rm g}=GM_{\rm bh}/c^2$.}
\cr
\end{tabular}
\label{t1}
\end{table}

\subsubsection{Orbital scenarios}

We consider two phenomenological scenarios based on the currently
available MAGIC observations. The first one is such that there is an
strict orbital periodicity of the neutrino emission, with the latter
being correlated with the $\gamma$-ray maximum measured by MAGIC
(Albert et al. 2005), say, within a period of about 10 days (the
active time) around phase 0.6 of the orbit. That is, the
$\gamma$-rays are produced during the time span where they have been
detected and there is no production of $\gamma$-rays at periastron.
A scenario where this could happen even in a hadronic description of
the $\gamma$-ray emission detected is briefly discussed below. A
second case considered here is one in which there is orbital
modulation of the $\gamma$-ray emission, i.e., $\gamma$-rays are
emitted along the orbit following the modulation imposed by the
target matter and the absorbing photon fields. The maximum of the
neutrino emission would then be naturally expected at periastron,
where the $\gamma$-ray production should also be maximal but
completely or nearly completely quenched by absorption. This
increases the active time along the orbit (we consider an active
time of 20 days) when significant neutrino emission proceeds.
As a benchmark, we show in Table 1 the results of the neutrino yield
for background and source events when there is no neutrino to photon
enhancement (i.e., photons are not absorbed). In this situation,
both for the cases of strict orbital periodicity and for the more
relaxed modulated signal, the background greatly dominates the
expected number of signal events. The question we pose then is how
much enhancement of the neutrino to photon ratio would still be
compatible with a non-detection of the system in a
km$^3$-observatory.

\subsubsection{Numerical results}

To fix first the numerics given by the current constraints on
neutrino emission, we consider that the neutrino flux follows the
MAGIC spectrum and is enhanced up to the level constrained by
AMANDA-II data (Ackerman 2006). In this situation, in the case of
periodicity, the signal events for energies below 1 TeV in a
detector such as ICECUBE during each orbital active time is 0.07
(with expected background, computed as explained above, of 0.12),
whereas it is again 0.07 (but with background 0.05) for energies
above 1 TeV.{ These quantities maybe slightly increased in
a more comprehensive treatment of the detector, what we explore below.} In the case of
modulation, these quantities are (at least) a factor of 2 larger,
given that the signal and background are proportional to the
integration time. In one year we have 14 LS I +61 303 orbits, such
that we expect about 2 signal events at all energies against 2.5 of
background in the case of periodicity; or 4 signal events at all
energies against a background of 5, in the case of modulation.
Detecting, in just one year, only the expected number of background
events or less, i.e., apparently no signal, in any of the cases
mentioned, would then discard the theoretically predicted signal at
the 68\% CL (Feldman \& Cousins 1998).

Instead of fixing a priori the level of enhancement of the neutrino
over the $\gamma$-ray flux, it can be consider as a variable upon
which constraints can be imposed from data.  The 95\% confidence
level interval for the mean number of 1-year signal events in the
case the number of detected events is equal to the expected
background is [0.00,5.75] and [0.00,6.26], for the periodicity and
modulation cases, respectively. The amplification factor between the
neutrino and photon fluxes is then limited to that which produces an
expected signal that is smaller than the upper end of these
intervals. We obtain $\sim$18 and $\sim$10, respectively. In one
year, then, a negative ICECUBE result for LS I +61 303 (i.e., a
detection of a number of neutrino events that is compatible with the
background expectation and are coming, within experimental
uncertainty, from the direction of LS I +61 303) would imply that
the amplification factor between the neutrino and $\gamma$-ray flux
is severely constrained. In addition, the level of constraint will
grow with additional observation time. Figure 1 describes this
improvement. As time goes by, the number of orbital periods that are
integrated is larger. This increases the background and signal
events expected from the theoretical assumptions. If just background
events continue to be measured,  Feldman \& Cousins' (1998) Table 4
and 5 gives the upper end of the 95\% CL interval for the mean
number of signal events that is still in agreement with the
experimental data. Assuming that the neutrino flux is enhanced from
that coming from the $\gamma$-ray spectrum measured, the maximum
neutrino to photon ratio that is compatible with the upper end of
that 95\% CL interval is obtained using Eq. (\ref{yellowsubmarine}).
This is what is plotted on the y-axis of the Figure 1. After a few
years of observations of just the background neutrino events, the
enhancement of the neutrino over the $\gamma$-ray flux allowed is so
low that theoretical models accounting for typical $\gamma$-$\gamma$
absorption (e.g., see Christiansen et al. 2006) cannot accommodate
it. Hadronic production of $\gamma$-rays would thus be indirectly
ruled out, with growing confidence. Alternatively, we can of course find the reverse result: the
predicted (lower limit) signal from LS I +61 303 in muon neutrinos
could lead to its discovery in ICECUBE.

\subsubsection{A more detailed treatment of the detector}

\begin{figure}[t]
\centering
\includegraphics[height=7cm, width=8.cm]{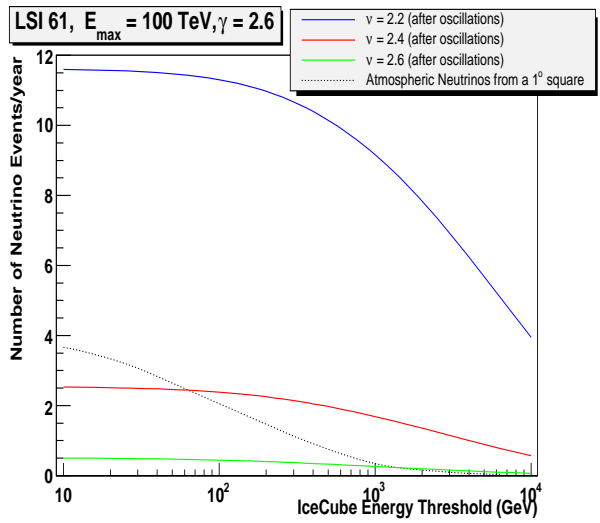}
\caption{ Event rate of neutrinos above an energy threshold, with a full account of ICECUBE effective area.}
\label{2}
\end{figure}

{ In Figure \ref{2} we show 
the event rate above an energy threshold, shown on the horizontal axis using a more detailed numerical description of the ICECUBE detector. Specifically, 
in this (alternative) calculation we obtain the neutrino flux using the bolometric method (Alvarez-Muniz \& Halzen 2002) that i) describes the spectra assuming the pionic origin of both gammas and neutrinos and ii) imposes equal energy in the flux of neutral and charged pions.We subsequently calculate the number of $\nu_\mu$ induced neutrino tracks in IceCube using the semianalytical calculation presented by 
Gonzalez-Garcia, Halzen \& Maltoni (2005)
with quality cuts on the IceCube data referred to as level 2 cuts by Ahrens et al. (2003). This anticipated performance of the IceCube detector is consistent with the initial data from the first 9 strings (see Achterberg et al. 2006).  When the spectra have equal high energy slopes, this method will reproduce our previous results, i.e., can be seen that these results agrees well with the result already quoted above, where an effective area of a kilometer squared has been assumed.    If we assume that the gamma-ray flux is steeper than the neutrino flux because of cascading, we get larger numbers for the neutrino predictions. These are also shown in the plot. }

\subsubsection{Electron neutrinos}

{ A concomitant way of looking
for neutrinos from LS I +61 303 would be the detection of electron
neutrinos (see, e.g., Section IIB of Anchordoqui \& Halzen 2005 for
details and formulae). The angular resolution of ICECUBE for the
detection of such neutrinos is not yet known but expected to several
degrees (say, several degrees) although with very large tails. That means
that for a fraction of the events, say half, it may be similar to
the AMANDA $\nu_\mu$ distribution: For those events with low
uncertainty in angular position, the background increase because of
the larger integration region would be compensated by the reduced
atmospheric $\nu_e$ background (a factor of $\sim$20 when compared
with $\nu_\mu$ at 1 TeV), thus probably producing a similar
confidence signal. The putative detection of LS I +61 303 in two
different channels would provide further confirmation of the
hadronic origin of the high energy radiation. Interesting to note in this case is that the
acceptance in showers is 4$\pi$ and ICECUBE 
would be able see the Southern micro-quasars as well}\\

\section{Discussion}

{ Note that for the former computations we have not made use of any particular model of the source or the nature of the system. Whatever they are, if the origin of the MAGIC-detected gamma-ray emission is hadronic, the neutrino to photon ratio can be constrained by ICECUBE, we have shown. We now briefly discuss our results in relation to published models of hadronic interpretations for LS I +61 303. Still, as we mentioned, a hadronic modelization of this system as a pulsar wind -- stellar wind interaction is not available, so we focus on microquasar models.}

A model of $\gamma$-ray and neutrino emission from microquasars
based on  hadronic interactions of relativistic protons in the jet
with stellar wind ions have been proposed by Romero et al. (2003)
and Torres et al. (2005). This kind of modelling has several free
parameters and embedded assumptions, and in order to test its
viability, it is of interest to compare the obtained results with
the report of the MAGIC detection of LS I +61 303.
The specific application of the quoted model to LS I +61 303 has
been developed by Romero et al. (2005) and Christiansen et al.
(2006). A prediction of this model as obtained in the previous works
is that the maximum of the $\gamma$-ray radiation is to be found at
periastron (even when considering in some detail the orbital
dependence of $\gamma$-$\gamma$ absorption, which is also maximal
there). This is not supported by MAGIC results, which have only
imposed upper limits at this orbital phase and found that the
maximum of the emission is instead shifted to the phases where the
radio and X-ray maxima are also located, after the periastron
passage. Is this fact enough to rule out { this particular hadronic interpretation}
(i.e., in the framework of the wind-jet interacting model) in
favor of leptonic ones (e.g., Bednarek 2006)?
In applying the hadronic modelling to LS I +61 303, and in order to
motivate a hadronic interpretation, Romero et al. (2005) have
emphasized the anti-correlation of the GeV and radio emission
reported by Massi 2004. This, however, overestimates the goodness
and orbital coverage of EGRET data on LS I +61 303. The typical
EGRET viewing period has a similar duration than the timescale that
is to be tested, i.e., the orbital period, which casts doubts upon
the resulting anti-correlation even if all other caveats (e.g.,
smallness of the data sample and errors for each of the lightcurve
points) were to be put aside. EGRET data can be used to claim
variability of the emission in monthly timescales (Torres et al.
2001, Nolan et al. 2003), and even that is not conclusive due to the
large error bars of each of the flux data points. Assigning a
precise orbital modulation at the GeV regime is clearly a task for
GLAST. { The former satellite will have a sensitivity about 30-50 times better than EGRET,
and in a matter of days will be able to detect the weakest of the EGRET sources. Recent simulations (Dubois 2006) show that GLAST will be able to follow the variabilty of LS I +61 303 in intra-orbit integrations, demonstrating the ability to detect the period of the system, should the GeV emission be correlated (or anti-correlated) with it.}
Massi (2004) proposed that the EGRET maximum is
correlated with the periastron of the system and Romero et al. and Christiansen et al.
used
this fact, requiring that the GeV flux at periastron is consistent
with the EGRET measurement, in order to fix a free phenomenological
parameter that accounts for some wind particles being unable to
diffuse into the microquasar jet. This parameter (so-called $f_p$)
linearly affects the computation of the $\gamma$-ray (and neutrino)
luminosity. $f_p$ was therefore chosen to be 0.1 all along the
orbit, this reduces one order of magnitude the otherwise predicted
GeV and TeV luminosity. { The parameter $f_p$ is ad-hoc, thus, it is not defined by any formula.}
It is likely that several effects can impede wind ions to enter
(diffuse) into the jet. Shock formation on the boundary layers or
partial alignment between jet and wind, are examples. { For instance, by
comparing the diffusion and convection timescales it can be seen
that efficient diffusion of ions into the jet is only possible when
the wind blows from the side (Torres et al. 2004). Thus, if the
wind-jet inclination changes with time, either by geometry or wind
dynamics, the efficiency in the diffusion of particles will also
change.} But these effects may not be constant along the orbit. In
fact, there is no reason to expect so when the wind velocity at the
impact with the jet, the accretion rate, the distance between stars,
and the wind density all significantly vary (order of magnitude)
between periastron (phase 0.23) and the $\gamma$-ray maximum (phase
$\sim 0.6$). In addition also the jet inclination with respect of
the stellar wind flow, particularly if the latter is not strictly
confined to the orbital plane, or if the angle of confinement change
randomly, may change with time. Just phenomenologically, then, if
the penetration factor $f_p$ changes significantly from 0 to 1 along
the orbit, even in a hadronic model, both strict periodicity (as
commented in the previous section), and thus, no radiation at
periastron, and a modulated scenario with emission all along the
orbit are possible. This is a conceptually new possibility for
hadronic models in microquasars, { which may even be useful for other cases should
the nature of LS I +61 303 finally reveals as jetless}, where radiation is quenched not
because of the existence of large stellar fields and thus an
increased opacity, but rather because of a modulation of the target
matter. { This reminds of collective stellar wind models
(Romero \& Torres 2003, Domingo-Santamar\'ia \& Torres 2006).} The fact that the phases at
which the $\gamma$-ray maxima occurs is located at the second
maximum of the accretion rate over the compact
object\footnote{Happening because the accretion rate is $\propto
v_{\rm rel}^{-3}$, where $v_{\rm rel}$ is the relative velocity
between the compact object and the stellar wind (Marti \& Paredes
1995, Gregory \& Neish 2002).} provides the needed target matter for
a hadronic interpretation of the $\gamma$-ray detection at this
orbital phase (provided the diffusion is efficient, $f_p \sim 1$),
whereas it reduces the resulting enhancement factor between the
neutrino and $\gamma$-ray flux (due to the reduced level of
$\gamma$-$\gamma$ opacity found far from the periastron of the
system). This implies that longer integration times in neutrino
telescopes such as ICECUBE are needed to reach ruling-out levels in
the strict periodicity scenario (consistent with Fig. 1). Additional
sources of opacity at periastron would decrease the level of
$\gamma$-ray flux, but increase the neutrino-to-photon ratio;
ICECUBE will be able to test this possibility.
Further MAGIC observations as well as ICECUBE results on LS I+61
303, a comparison of these  with the results of the previous
section, and a more detailed implementation of the conceptual
ideas about suppression of target matter at periastron, will help
further constrain, or finally rule out, a hadronic interpretation
for this system.

\section*{Acknowledgements}

DFT thanks Luis Anchordoqui and Eva Domingo-Santamar\'ia 
and his colleagues of the
MAGIC collaboration, especially W. Bednarek, V. Bosch-Ramon, J.
Cortina, J. M. Paredes, M. Ribo, J. Rico, and N. Sidro, for
discussions on the $\gamma$-ray detection from LS I +61 303. This research
been supported by Ministerio de Educaci\'on y Ciencia (Spain) 
under grant AYA-2006-00530, and by the Guggenheim Foundation.


\begin{thebibliography}{10}

\bibitem{Achterberg:2006pw}
  Achterberg  A. et al. (IceCube Collaboration),
  ``Contributions to 2nd TeV particle astrophysics conference (TeV PA II)
  Madison Wisconsin - 28-31 August 2006,''
  arXiv:astro-ph/0611597.
  %%CITATION = ASTRO-PH 0611597;%%

\bibitem{} Ackermann M. et al. 2005, Phys. Rev. D 71, 077102

\bibitem{} Ackermann M. 2006, for the AMANDA Collaboration, talk presented at the meeting on
``The Multi-Messenger Approach To High-Energy Gamma-Ray Sources"
held in Barcelona, July 4-7, 2006. To be published in the
proceedings. The talk is electronically available at
http://www.am.ub.es/bcn06/

\bibitem{} Aharonian F. et al. (HESS Collaboration) 2005, Science 309, 746

\bibitem{} Aharonian et al. (HESS Collaboration) 2006,
 astro-ph/0611813, A\&A, in press 

\bibitem{} Aharonian et al. (HESS Collaboration) 2006b,
 astro-ph/0612495, ApJ, in press 

\bibitem{}  Aharonian F., Anchordoqui L. A., Khangulyan D., \&
Montaruli T. 2005, astro-ph/0508658

\bibitem{} Ahrens J. et al. (ICECUBE Collaboration) 2004, Astropart. Phys. 20, 507
%%CITATION = ASTRO-PH 0305196;%%

\bibitem{} Albert J. et al. (MAGIC Collaboration) 2006, Science 312,
1771
  %%CITATION = ASTRO-PH 0605549;%%

\bibitem{} Alvarez-Muniz J. \& Halzen F. 2002, ApJ 576, L33

\bibitem{}Anchordoqui L. A., Torres D. F., McCauley T. P.,
Romero G. E. \& Aharonian F. A. 2003, ApJ 589, 48

\bibitem{} Anchordoqui L. A., \& Halzen F. 2005, hep-ph/0510389

\bibitem{} Bednarek W. 2006, Mon. Not. R. Astron Soc. 368, 579

\bibitem{} Bosch-Ramon V., Aharonian F. A., \& Paredes J. M. 2005,
A\&A 432, 609

\bibitem{} Bosch-Ramon V., et al. 2006, A\&A 466, 1081
%%CITATION = ASTRO-PH 0601238;

\bibitem{}  Casares J., et al. 2005, Mon. Not. R. Astron. Soc. 360,
1105.

\bibitem{} Cavassinni V., Grasso D., \& Maccione L. 2006, astro-ph
0604004

\bibitem{} Christiansen H. R., Orellana M., Romero G. E. 2006,
to appear in Phys. Rev. D., astro-ph/0509214

\bibitem{}  Costantini M. L. \& Vissani F. 2005, Astropart. Phys. 23, 477

\bibitem{} Dhawan V. et al. 2006, reported in the Conference on Microquasars, Como, Italy, 18-22 September, to appear in the Proceedings of Science.

\bibitem{} Distefano C., Guetta D., Waxman E. \& Levinson A. 2002, ApJ 575, 378

\bibitem{} Distefano C. et al. (NEMO Collaboration) 2006, astro-ph/0608514, in the Proceeding of "The multi messenger approach to high energy gamma ray sources", Barcelona, June 2006, Kluwer Academics, in press (Paredes J.M., Reimer O., and Torres D. F. Editors)


\bibitem{} Domingo-Santamar\'ia E. \& Torres D. F. 2005, A\&A 444,
403
  %%CITATION = ASTRO-PH 0506240;%%

\bibitem{} Domingo-Santamar\'ia E. \& Torres D. F. 2006, A\&A 448,
613
  %%CITATION = ASTRO-PH 0510769;%%

\bibitem{} Dubois R., 2006, reported in the Conference on Microquasars, Como, Italy, 18-22 September, to appear in the Proceedings of Science. SLAC-PUB-12174 (November 2006)


\bibitem{} Feldman G. J. \& Cousins R. D. 1998, Phys.Rev. D57, 3873

\bibitem{}  Frail D. A. \& Hjellming, R. M. 1991,Astron. J. 101,
2126

\bibitem{}  Goldoni P. \& Mereghetti, S. 1995, Astron. Astrophys. 299, 751


\bibitem{Gonzalez-Garcia:2005xw}
  Gonzalez-Garcia M. C., Halzen F, \& Maltoni M. 2005,
  %``Physics reach of high-energy and high-statistics Icecube
  %atmospheric neutrino data,''
  Phys.\ Rev.\ D { 71}, 093010 
  %%CITATION = HEP-PH 0502223;%%

\bibitem{} Gregory P. C. 2002, Astrophys. J. 575, 427



\bibitem{}
Gaisser T.K., Halzen F. and Stanev T. 1995,
%``Particle astrophysics with high-energy neutrinos,''
Phys.\ Rept.\  { 258}, 173 [Erratum-ibid.\  { 271}, 355 (1996)].
%[arXiv:hep-ph/9410384].
%%CITATION = HEP-PH 9410384;%%

\bibitem{} Gregory P. C., \& Neish C. 2002, ApJ 580, 1133

\bibitem{} Halzen F., astro-ph/0602132, to appear in Eur.Phys.J. C

\bibitem{} Halzen F., \& Hooper D. 2005, Astropart. Phys. 23, 537

\bibitem{} Harrison, F. A., Ray, P. S., Leahy, D. A., Waltman, E. B. \& Pooley,
G.G. 2000, Astrophys. J. 528, 454

\bibitem{} Hartman R. C. et al. 1999, Astrophys. J. Supl. 123, 79

\bibitem{} Heinz S., \& Sunayev R. 2002, A\&A 390, 751

\bibitem{} Hutchings J. B. \& Crampton D. 1981, Pub. Astron. Soc. Pacific 93, 486





\bibitem{} Kawachi A. et al. (CANGAROO Collaboration) 2004, ApJ 607, 949

\bibitem{} Kirk J. G., Ball L., \& Skjaeraasen O. 1999, Astroparticle 
Physics 10, 31 





\bibitem{} Levinson A. \& Blanford R. 1996, ApJ 456, L29

\bibitem{} Levinson A. \& Waxman E. 2001, Phys. Rev. Lett. 87, 171101


\bibitem{} Levinson A. 2006, astro-ph/0611521, to appear in Mod. Phys. Lett. A.


%\bibitem{Lipari:hd}
\bibitem{} Lipari P. 1993,
%``Lepton Spectra In The Earth's Atmosphere,''
Astropart.\ Phys.\  { 1}, 195.
%%CITATION = APHYE,1,195;%%

\bibitem{} Malkov M. A. and OÕC Drury L. 2001, Rept. Prog. Phys. 64, 429 


\bibitem{} Marti J., \& Paredes J. M. 1995, A\&A 298, 151

\bibitem{} Massi M. 2004, A\&A 422, 267

\bibitem{} Massi M., Rib\'o M., Paredes J. M., Peracaula M. \& Estalella
R. 2001, A\&A 376, 217

\bibitem{} Massi M. et al. 2004, Astron. Astrophys. 414, L1

\bibitem{}Migliari, S., Fender, R. \& M\'endez, M. 2002, Science, 297, 1673

\bibitem{} Neronov A. \& Chernyakova M. 2006, astro-ph/0610139, in the Proceeding of "The multi messenger approach to high energy gamma ray sources", Barcelona, June 2006, Kluwer Academics, in press (Paredes J.M., Reimer O., and Torres D. F. Editors)


\bibitem{} Nolan P., Tompkins W. F., Grenier I. A., \& Michelson P. 2003, ApJ 597, 615


\bibitem{} Romero G. E. Kaufman-Bernado M. M., Combi J., \& Torres D. F. 2001, A\&A 376, 599
%%CITATION = ASTRO-PH 0107411;

\bibitem{} Romero G. E. \& Torres D. F. 2003, ApJ 586, L33
  %%CITATION = ASTRO-PH 0302149;%%

\bibitem{} Romero G. E., Torres D. F., Kaufman Bernado M. M., \&
Mirabel I. F. 2003, A\&A 410, L1
  %%CITATION = ASTRO-PH 0309123;%%


\bibitem{} Romero G. E., Christiansen H. R., \& Orellana M. 2005,
ApJ 632, 1093

\bibitem{} Romero G. E. \& Torres D. F 2003, ApJ 586, L33
%%CITATION = ASTRO-PH 0302149;%%

\bibitem{} Taylor A. R., Young G., Peracaula M., Kenny H. T. \& Gregory P. C. 1996, Astron. Astrophys. 305, 817

\bibitem{}Torres D. F., et al. 2001, A\&A, 370,
468
  %%CITATION = ASTRO-PH 0007464;%%


\bibitem{} Torres D. F., et al. 2003, Phys. Rept. 382, 303
%%CITATION = ASTRO-PH 0209565;

\bibitem{} Torres D. F. 2004, ApJ 617, 966
  %%CITATION = ASTRO-PH 0407240;%%

\bibitem{} Torres D. F., Domingo-Santamar\'ia E., \& Romero G. E.
2004, ApJ 601, L75
  %%CITATION = ASTRO-PH 0312128;%%

\bibitem{} Torres D. F., Romero G. E., \& Mirabel I. F. 2005,
Chin. J. Astron. Astrophys. Suppl. 5, 183 (astro-ph/0407494)
  %%CITATION = ASTRO-PH 0407494;%%

\bibitem{Volkova:sw}
Volkova L.V. 1980,
%``Energy Spectra And Angular Distributions Of Atmospheric Neutrinos,''
Sov.\ J.\ Nucl.\ Phys.\  { 31}, 784  [Yad.\ Fiz.\  { 31}, 1510
(1980)].
%%CITATION = SJNCA,31,784;%%

\end{thebibliography}
\end{document}